\documentclass[twoside,twocolumn,9pt]{article}
\usepackage{extsizes}
\usepackage{amsmath,amssymb}
\usepackage[super,sort&compress,comma]{natbib}
\usepackage[version=3]{mhchem}
\usepackage[left=1.5cm, right=1.5cm, top=1.785cm, bottom=2.0cm]{geometry}
\usepackage{balance}
\usepackage{mathptmx}
\usepackage{sectsty}
\usepackage{graphicx}
\usepackage{lastpage}
\usepackage[format=plain,justification=justified,singlelinecheck=false,font={stretch=1.125,small,sf},labelfont=bf,labelsep=space]{caption}
\usepackage{float}
\usepackage{fancyhdr}
\usepackage{fnpos}
\usepackage[english]{babel}
\addto{\captionsenglish}{%
  
}
\usepackage{array}
\usepackage{droidsans}
\usepackage{charter}
\usepackage[T1]{fontenc}
\usepackage[usenames,dvipsnames]{xcolor}
\usepackage{setspace}
\usepackage[compact]{titlesec}
\usepackage{hyperref}

\definecolor{cream}{RGB}{222,217,201}

\begin{document}

\pagestyle{fancy}
\thispagestyle{plain}
\fancypagestyle{plain}{
\renewcommand{\headrulewidth}{0pt}
}

\makeFNbottom
\makeatletter
\renewcommand\LARGE{\@setfontsize\LARGE{15pt}{17}}
\renewcommand\Large{\@setfontsize\Large{12pt}{14}}
\renewcommand\large{\@setfontsize\large{10pt}{12}}
\renewcommand\footnotesize{\@setfontsize\footnotesize{7pt}{10}}
\makeatother

\renewcommand{\thefootnote}{\fnsymbol{footnote}}
\renewcommand\footnoterule{\vspace*{1pt}%
\color{cream}\hrule width 3.5in height 0.4pt \color{black}\vspace*{5pt}}
\setcounter{secnumdepth}{5}

\makeatletter
\renewcommand\@biblabel[1]{#1}
\renewcommand\@makefntext[1]%
{\noindent\makebox[0pt][r]{\@thefnmark\,}#1}
\makeatother
\renewcommand{\figurename}{\small{Fig.}~}
\sectionfont{\sffamily\Large}
\subsectionfont{\normalsize}
\subsubsectionfont{\bf}
\setstretch{1.125} 
\setlength{\skip\footins}{0.8cm}
\setlength{\footnotesep}{0.25cm}
\setlength{\jot}{10pt}
\titlespacing*{\section}{0pt}{4pt}{4pt}
\titlespacing*{\subsection}{0pt}{15pt}{1pt}

\fancyfoot{}
\fancyfoot[RO]{\footnotesize{\sffamily{1--\pageref{LastPage} ~\textbar  \hspace{2pt}\thepage}}}
\fancyfoot[LE]{\footnotesize{\sffamily{\thepage~\textbar\hspace{3.45cm} 1--\pageref{LastPage}}}}
\fancyhead{}
\renewcommand{\headrulewidth}{0pt}
\renewcommand{\footrulewidth}{0pt}
\setlength{\arrayrulewidth}{1pt}
\setlength{\columnsep}{6.5mm}
\setlength\bibsep{1pt}

\makeatletter
\newlength{\figrulesep}
\setlength{\figrulesep}{0.5\textfloatsep}

\newcommand{\topfigrule}{\vspace*{-1pt}%
\noindent{\color{cream}\rule[-\figrulesep]{\columnwidth}{1.5pt}} }

\newcommand{\botfigrule}{\vspace*{-2pt}%
\noindent{\color{cream}\rule[\figrulesep]{\columnwidth}{1.5pt}} }

\newcommand{\dblfigrule}{\vspace*{-1pt}%
\noindent{\color{cream}\rule[-\figrulesep]{\textwidth}{1.5pt}} }

\makeatother

\twocolumn[
  \begin{@twocolumnfalse}
{
}\par
\vspace{1em}
\sffamily
\begin{tabular}{m{4.5cm} p{13.5cm} }

& \noindent\LARGE{\textbf{Electrical conductivity of random metallic nanowire networks: An analytical consideration along with computer simulation}} \\
\vspace{0.3cm} & \vspace{0.3cm} \\

 & \noindent\large{Yuri Yu. Tarasevich,$^{\ast}$\textit{$^{a}$} Irina V. Vodolazskaya,\textit{$^{a}$} and Andrei V. Eserkepov\textit{$^{a}$}} \\

& \noindent\normalsize{
We have proposed an analytical model for the electrical conductivity in random, metallic, nanowire networks. We have mimicked such random nanowire networks as random resistor networks (RRN) produced by the homogeneous, isotropic, and random deposition of conductive zero-width sticks onto an insulating substrate. We studied the electrical conductivity of these RRNs using a mean-field approximation. An analytical dependency of the electrical conductivity on the main physical parameters (the number density and electrical resistances of these wires and of the junctions between them) has been derived. Computer simulations have been performed to validate our theoretical predictions. We computed the electrical conductivity of the RRNs against the number density of the conductive fillers for the junction-resistance-dominated case and for the case where the wire resistance and the junction resistance were equal. The results of the computations were compared with this mean-field approximation. Our computations  demonstrated that our analytical expression correctly predicts the electrical conductivity across a wide range of number densities.} \\
\end{tabular}

 \end{@twocolumnfalse} \vspace{0.6cm}

  ]

\renewcommand*\rmdefault{bch}\normalfont\upshape
\rmfamily
\section*{}
\vspace{-1cm}


\footnotetext{\textit{$^{a}$~Laboratory of Mathematical Modeling, Astrakhan State University, Astrakhan 414056, Russia}}
\footnotetext{\textit{$^{\ast}$~ E-mail: tarasevich@asu.edu.ru }}


\section{Introduction}\label{sec:intro}

Numerous modern optoelectronic devices such as  touch-screens, heaters, and solar cells contain transparent electrodes as their important components.\cite{Hecht2011AM,McCoul2016AEM,Sannicolo2016Small,Gao2016AdvPhys,ZhangChemRev2020,Patil2020} A transparent, poorly conductive film containing randomly distributed highly conductive elongated fillers such as nanowires, nanotubes, or  nanorods presents an example of one of the possible kinds of transparent electrodes.\cite{Nam2016N,Ackermann2016SR,Callaghan2016PCCP,McCoul2016AEM,Zhang2017JMM,Hicks2018JAP,Glier2020,Balberg2020}
Among these, flexible, transparent, conductive materials based on metal nanowire networks (e.g., \ce{Cu}, \ce{Ag}, \ce{Au}) are especially widely used.\cite{Langley2013,Ye2014,Liu2019} Numerous efforts are currently underway to identify the main factors affecting the effective electrical conductivity, transparency, and haze of such transparent electrodes.\cite{Wang2008MSE,Mutiso2015PPS}

There are different ways to describe the electrical conductivity of random nanowire networks.
On the one hand, a set of linear equations from Kirchhoff's rules and Ohm's laws can be written for any given nanowire network; although, this set, as a rule, is huge, it may, in principle, be  solved numerically for obtaining potentials and currents.\cite{Kirkpatrick1971PRL,Kirkpatrick1973RMP,Li2007JPhA,Benda2019,Kim2018JAP,Kim2020JCPC} A significant drawback of such numerical methods is that they present the electrical conductivity as plots or tables rather than as analytical expressions connecting the electrical conductivity with the key physical parameters.

On the other hand, the electrical conductivity of random nanowire networks may be treated theoretically when using a range of different approaches, e.g., percolation theory,\cite{Zezelj2012PRB} effective medium theory,\cite{Callaghan2016PCCP} geometrical consideration,\cite{Kumar2016JAP,Kumar2017JAP,Gupta2017} or mean-field approximation.\cite{Forro2018ACSN,Tarasevich2021arXiv} However, analytical formulas often contain some adjustable parameters, and their derivation is not always based on rigorously proven assumptions.

Thus, the sheet resistance, $R_\Box$, of dense networks of randomly placed widthless sticks can be calculated as
\begin{equation}\label{eq:RsheetZS}
   R_\Box = \frac{b n^{t-1} R_\text{s} + (n + n_\text{c})^{t-2}R_\text{j}}{a\left[(n-n_\text{c})^t + c(L/l)^{-t/\nu}\right]},
\end{equation}
where $L$ is the linear system size, $l$ is the length of the stick, $R_\text{s}$ is the resistance of the stick, $R_\text{j}$ is the junction resistance, $n$ is the number density of the conductive sticks (number of sticks per unit area), $n_\text{c}$ is the percolation threshold, $a$, $b$, and $c$ are adjustable parameters, $\nu$ is the correlation-length exponent, and $t$ is the conductivity exponent.\cite{Zezelj2012PRB} This relation is based on the assumption that the electrical conductivity of the system can be modelled as an equivalent serial conductance with $n$ sticks in parallel and $n^2$ junctions in parallel, when the number density of the sticks is well above the percolation threshold $n \gg n_\text{c}$.

Based on a geometrical consideration of a thin film of randomly deposited conductive wires, a formula for the sheet resistance has been proposed\cite{Kumar2017JAP}
\begin{equation}\label{eq:KumarR1}
  R_\Box = \frac{\pi}{2\sqrt{N_E}} \left( \frac{4 l \rho}{\pi D^2} + \frac{R_\text{j}}{d}\right),
\end{equation}
where $\rho$ is the electrical resistivity of the wire material, and $D$ is the wire diameter,
$$
N_E = nl^2\left[Cnl^2 + \exp\left(-Cnl^2\right) - 1\right],
$$
$$
d = \frac{1 - \exp\left(-Cnl^2\right)}{ Cnl^2} - \exp\left(-Cnl^2\right).$$
Here,
$C = 2/\pi$. We have written~\eqref{eq:KumarR1} for an arbitrary value of $l$, while, in the original article,\cite{Kumar2017JAP} the wire length, $l$, was assumed to be unity. The sheet resistance can be written as follows
\begin{equation}\label{eq:KumarR}
R_\Box =  \frac{\pi}{2 d \sqrt{N_E}} \left( d R_\text{s} + R_\text{j} \right).
\end{equation}
Note, that $R_\Box $ tends to the constant value $R_\text{j} /\sqrt{C}$, when $R_\text{s} \ll R_\text{j}$, while $n \gg 1$. Such a behaviour is hardly reasonable.

An alternative formula has been proposed using a mean-field approximation (MFA).\cite{Forro2018ACSN} Rather than study all the conductors in a system, an MFA involves considering a single conductor, placed in the mean  electric field that is produced by all the other conductors.
\begin{equation}\label{eq:Forro}
  R_\Box  = \frac{R_\text{s}}{n l^2}\left[ \frac{r_\text{m}}{2} - \sqrt{ \frac{R_\text{j} r_\text{m}}{ R_\text{s} C n l^2}}\tanh \left(\sqrt{\frac{R_\text{s} C n l^2 r_\text{m}}{4 R_\text{j}}}\right)\right]^{-1}.
\end{equation}
Here, $n_\text{a} = nCl^2/2$ is the effective number of contacts. The authors argued that, on average, only half of the contacts carry current, since current mainly flows through those contacts where the end of a wire is connected to the beginning of another.
$$
r_\text{m} = \frac{n_\text{a} - 1 +R_\text{j}\left( R_\text{j} + \frac{R_\text{s}}{n_\text{a} + 1} \right)^{-1}}{n_\text{a} + 1}
$$
is the effective wire length postulated without any reasonable speculations.
(We have changed the original notation to provide uniformity throughout this text.)

The effect of junction resistance on the conductivity of nanowire- and nanotube-based conductive networks has been analyzed.\cite{Zezelj2012PRB,Rocha2015,Ponzoni2019APL,Fata2020JAP} Typically, in untreated nanowire-based networks, the wire-to-wire junction resistances dominate over the resistance of the nanowires, themselves.\cite{Bellew2015,Manning2020}

Based on an analysis of such experimental data, a simple relation has recently been proposed
$$
R_\Box \approx \alpha\left( R_n + R_\text{j}\right),
$$
where $ R_n $ is the average value of the electrical resistance between two contacts, while $\alpha$ is an adjustable parameter.\cite{Ponzoni2019APL} A similar linear relation additionally accounting for the metallic electrode/nanowire contact resistance has been obtained\cite{Benda2019} by applying the Buckingham $\Pi$-theorem.\cite{Buckingham1914}

The goal of the present work is an investigation of the electrical properties of 2D disordered systems involving an insulating host matrix and conductive rod-like fillers (zero-width sticks). The number density of the conductive fillers ranges from the percolation threshold, $n_\text{c}$, to $\approx 20 n_\text{c}$. The system under consideration is treated as a random resistor network (RRN). We use a particular kind of MFA.
Two cases represent our focus, viz.,
\begin{enumerate}
  \item Unwelded wires. The junction resistance dominates over the wire resistance ($R_\text{j} \gg R_\text{s}$).
  \item Welded wires (see Refs.~\citenum{Kang2018,Ding2020} for a review of welding techniques). Here, the junction resistance and the wire resistance are of the same order (in our study, $R_\text{j} = R_\text{s}$).
\end{enumerate}

The rest of the paper is constructed as follows. Section~\ref{sec:methods} describes some technical details of the simulation (Sec.~\ref{subsec:simul}) and our variant of the MFA (Sec.~\ref{subsec:MFA}). In Section~\ref{sec:results}, we presents our main results and discuss their reliability. Section~\ref{sec:concl} summarizes the main results and suggests possible directions for further generalization of the model.

\section{Methods}\label{sec:methods}
\subsection{Simulation}\label{subsec:simul}

Zero-width sticks of length $l$ were placed within a square domain $\mathcal{D}$ of size $L \times L$ ($L>l$) with periodic boundary conditions. The centres of the sticks were assumed to be independent and identically distributed (i.i.d.) within  $\mathcal{D} \in \mathbb{R}^2$, i.e., $x,y \in [0;L]$, where $(x,y)$ are the coordinates of the centre of the stick under consideration. Their orientations were assumed to be equiprobable. In such a way, a homogeneous and isotropic network was produced. In our simulations, without loss of generality, sticks of unit length were used ($l = 1$). We used a system of size $L=32$.

The Union--Find algorithm~\cite{Newman2000PRL,Newman2001PRE} modified for continuous systems~\cite{Li2009PRE,Mertens2012PRE} was applied to detect the percolation cluster. When a percolation  cluster was found, an adjacency matrix was formed for it.  All other clusters were ignored since they cannot contribute to the electrical conductivity. With this adjacency matrix in hand, an RRN could be constructed. Each stick was treated as a resistor with an electrical resistance $R_\text{s}$. Each contact between any two sticks was treated as a junction with an  electrical resistance $R_\text{j}$. Accordingly, a segment of stick between two nearest contacts corresponds to  a resistor with an electrical resistance, $R_\text{s} l_k/l$, where $l_k$ is the distance between these two contacts. In such a way, an RRN consisting of two kinds of resistors was considered.

Kirchhoff's current law was used for each junction between two resistors of any kind, while Ohm’s law was applied to each resistor between any two junctions. The resulting set of linear equations with a sparse matrix was solved using \emph{Eigen}~\cite{eigenweb}, a C++ template library for linear algebra. Since only square samples were considered, the electrical conductivity is simply the inverse of the sheet resistance, i.e., $\sigma = R_\Box^{-1}$.

The computer experiments were repeated 100 times for each value of the number density. The error bars in the figures correspond to the standard deviation of the mean. When not shown explicitly, they are of the order of the marker size.

\subsection{Mean-field approach}\label{subsec:MFA}
\begin{figure}[!htb]
  \centering
  \includegraphics[width=0.85\columnwidth]{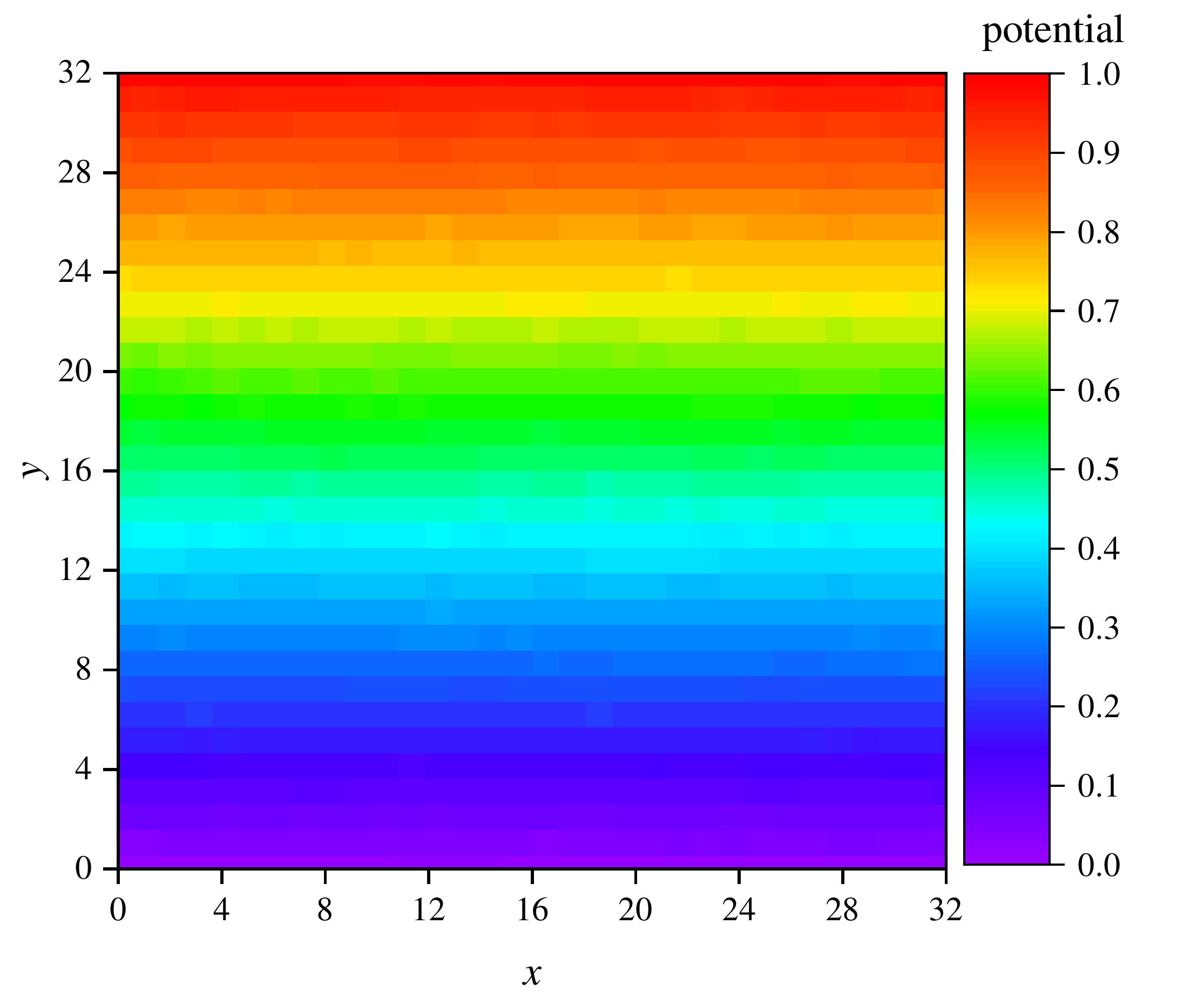}\\
  \caption{Example of simulated potential distribution in the system under consideration $L=32$ when $n - n_c = 45$, where $n_c$ is the percolation threshold; for this particular system $n_c = 5.7813$, $R_\text{s}=R_\text{j}$. The entirely domain is divided into unit cells (squares); the potentials of all contacts within each cell were averaged (about 1600 contacts in each cell). \label{fig:potent}}
\end{figure}

In dense systems, the voltage drop along the system is close to linear.\cite{Bergin2012Nanoscale,Khanarian2013JAP,Sannicolo2018,Forro2018ACSN,Papanastasiou2021,Charvin2021}  This statement is also confirmed by our simulation (Fig.~\ref{fig:potent}). Instead of considering the entire ensemble of conductors, here, we consider a single conductor located in the average electrostatic field created by all the other conductors. When a potential difference, $V_0$, is applied to two opposite borders of the system,  this field is
\begin{equation}\label{eq:E}
E = \frac{V_0}{L}.
\end{equation}

\subsubsection{Common consideration}
\begin{figure}[!htb]
  \centering
  \includegraphics[width=0.45\columnwidth]{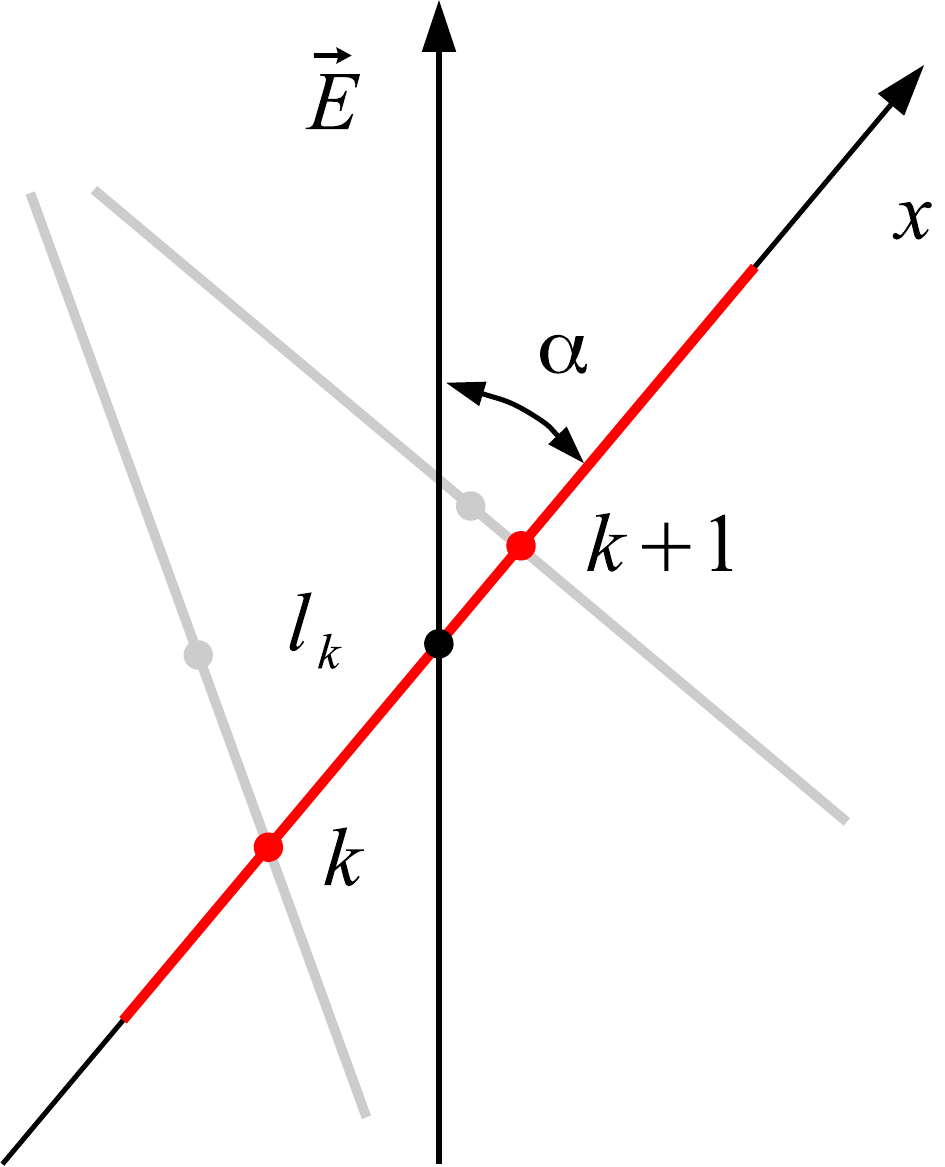}\\
  \caption{A linear conductor of length $l$ in an external electric field (red). Only two contacts are indicated. Their numbers are $k$ and $k+1$. The distance between these two contacts is $l_k$. The contact numbers go from 1 to $N_j$, while the numbers of segments, into which the contacts divide the conductor,  run from 0 to $N_j$.}\label{fig:contacts}
\end{figure}

Let there be a linear conductor of length $l$ and resistance $R_\text{s}$, which is immersed in an external homogeneous electric field $\mathbf{E}$. The angle between this conductor and the field is $\alpha$ (Fig.~\ref{fig:contacts}). Let there be $N_j$ point contacts evenly distributed along the conductor. These contacts provide some leakage conductivity. The resistance of each contact (junction) is $R_\text{j}$,  while the distance between any two nearest contacts is
$$
l_k = \frac{l}{N_j + 1}.
$$

According to Ohm's law, the potential drop between the two nearest contacts $k$ and $k + 1$ equals
\begin{equation}\label{eq:u}
  u_{k+1} - u_k + \frac{l_k}{l} R_\text{s} i_k = 0,
\end{equation}
where $i_k$ is the electric current between the contacts $k$ and $k + 1$.

The change in the electric current when passing through contact number $k$ is due to a loss of charge in this contact
\begin{equation}\label{eq:i}
i_k - i_{k-1} + \frac{u_k - V_k}{R_\text{j}}  = 0,
\end{equation}
where $V_k$ is the potential of the external electrical field at the point where the contact of number $k$ is located.  Equations~\eqref{eq:u} and~\eqref{eq:i} can be combined
\begin{equation}\label{eq:inhomoi1}
   i_{k+1} - \mu i_k  + i_{k-1} = -\frac{E l \cos \alpha}{R_\text{j}(N_j + 1)},
\end{equation}
where
$$
\mu = 2 + \frac{\Delta}{N_j + 1}, \quad \Delta = \frac{R_\text{s}}{R_\text{j}}.
$$
The solution of this linear recurrence with constant coefficients
\begin{equation}\label{eq:isolution}
  i_k(\alpha,N_j) = \frac{E l \cos \alpha}{R_\text{s}} \left[ 1 + \frac{ \left(\lambda_2 ^{ N_j} - 1\right) \lambda_1^k  -  \left( \lambda_1 ^{ N_j} - 1 \right) \lambda_2^k }{ \lambda_1 ^{N_j}- \lambda_2^{N_j}}\right]
\end{equation}
describes the electric current in the $k$-th segment of the linear conductor having exactly $N_j$ evenly distributed contacts, and which is located at an angle $\alpha$ to the external electric field. Here,
\begin{equation}\label{eq:lambda}
\lambda_{1,2} = \frac{\mu \pm \sqrt{\mu^2 - 4}}{2}.
\end{equation}

\begin{figure}[!htb]
  \centering
  \includegraphics[width=0.85\columnwidth]{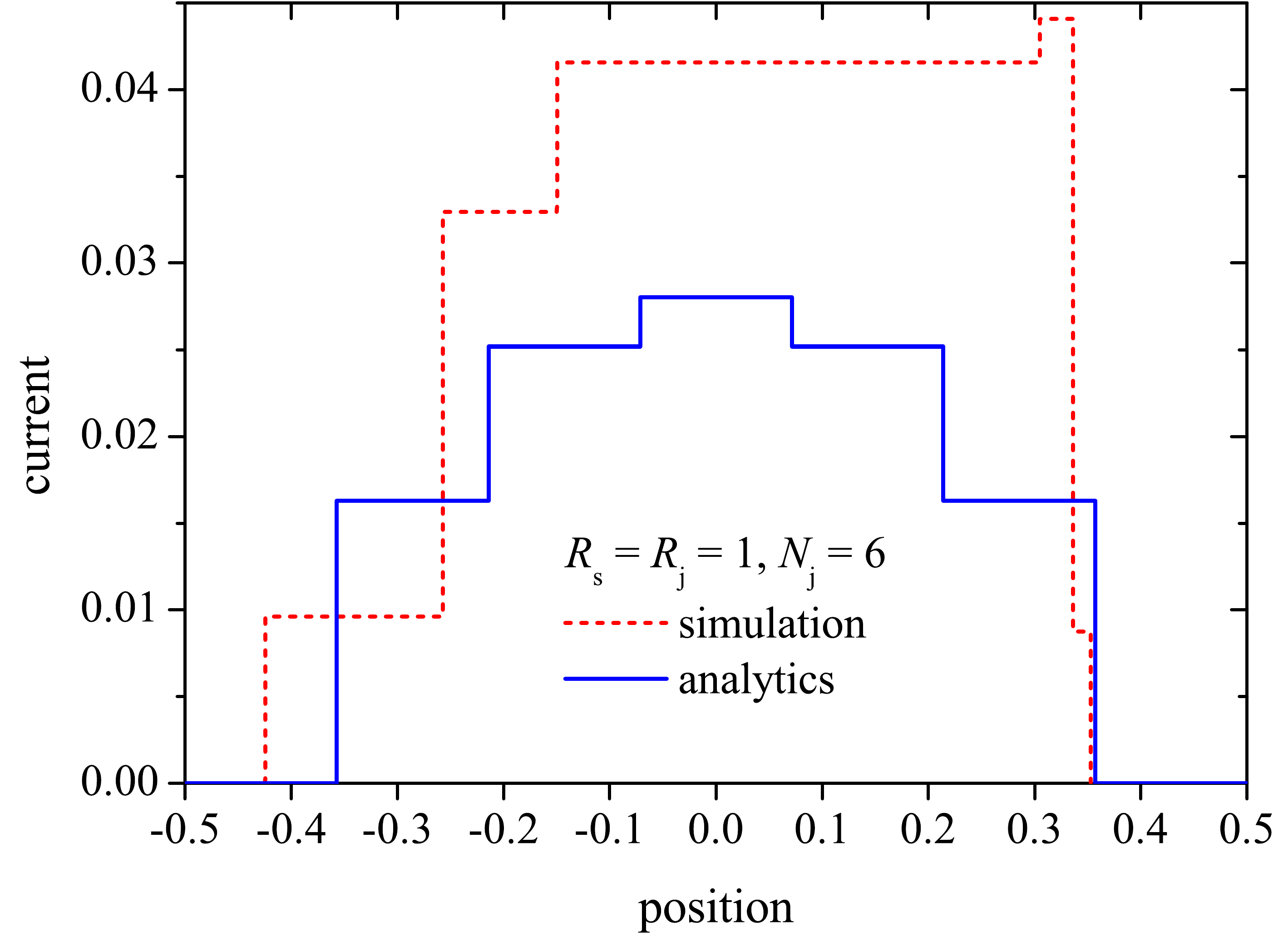}
  \caption{Example of current distribution calculations in the segments of one randomly selected conductor with 6 contacts (dashed line) and the results of calculations using  formula~\eqref{eq:isolution} (solid line). $R_\text{s}=R_\text{j}=1$, $n = 3\pi$, $\alpha = -1.32751$.}\label{fig:current1compare}
\end{figure}

Figure~\ref{fig:current1compare} shows an example of calculating the distribution of the currents in the segments of one particular randomly selected conductor with 6 contacts (dashed line) and the results of calculations using formula~\eqref{eq:isolution} (solid line) when $R_\text{s} = R_\text{j}$, $n = 3\pi$, and $\alpha = -1.32751$.

The average electric current in a conductor with $N_j$  contacts and which is located at an angle $\alpha$ to the external electric field is equal to
$$
  \langle i(\alpha, N_j)\rangle = \frac{E l \cos \alpha}{ R_\text{s}}\left[ 1 + \frac{1}{N_j+1}\sum_{k=0}^{N_j}\frac{ \left(\lambda_2 ^{ N_j} - 1\right) \lambda_1^k  -  \left( \lambda_1 ^{ N_j} - 1 \right) \lambda_2^k }{ \lambda_1 ^{N_j}- \lambda_2^{N_j}}\right].
$$
The sum can be easily calculated, which leads to
$$
  \langle i\left(\alpha,N_j\right)\rangle = \frac{E l \cos \alpha}{ R_\text{s}}
  \left( 1-\frac{2}{\Delta}\frac{\lambda_1^{ N_j  +1} - \lambda_2^{ N_j  +1}
+\lambda_2^{ N_j }-\lambda_1^{ N_j }
 + \lambda_2-  \lambda_1}{\lambda_1^{ N_j } - \lambda_2^{ N_j }}\right).
$$

Now, after consideration of an abstract system,  we can turn to our system of interest described in Section~\ref{subsec:simul}.  Since all orientations of a stick are equiprobable, the number of sticks intersecting a line of length $L$ perpendicular to the field is $ n l L \cos \alpha $. The total electric current in the sticks of all orientations through a cross section of the system is
$$
\frac{n l L }{\pi} \int_{-\pi/2}^{\pi/2} \left\langle i\left(\alpha,\left\langle N_j\right\rangle\right) \right\rangle \cos \alpha \, d\alpha,
$$
where $\left\langle N_j\right\rangle$ is the average number of contacts per a stick. In the system under consideration $\left\langle N_j\right\rangle = 2 n l^2 /\pi$.
Hence,
\begin{multline}\label{eq:mean-i-vs-Nj}
\langle i \rangle =\\  \frac{ E n l^2 L  }{ 2 R_\text{s}}
\left( 1-\frac{2}{\Delta}\frac{\lambda_1^{\left\langle N_j\right\rangle  +1} - \lambda_2^{ \left\langle N_j \right\rangle  +1}
+\lambda_2^{ \left\langle N_j\right\rangle }-\lambda_1^{ N_j }
 + \lambda_2-  \lambda_1}{\lambda_1^{ \left\langle N_j\right\rangle } - \lambda_2^{ \left\langle N_j\right\rangle }}\right).
\end{multline}
According to Ohm's law, the electrical conductivity is
\begin{multline}\label{eq:MFAsigma-discr}
\sigma = \\ \frac{\langle N_j \rangle }{ 2 C R_\text{s}}
\left(
1 -
\frac{2}{\Delta}\frac{\lambda_1^{\langle N_j \rangle +1} - \lambda_2^{\langle N_j \rangle +1}
+\lambda_2^{\langle N_j \rangle}-\lambda_1^{\langle N_j \rangle}
 + \lambda_2-  \lambda_1}{\lambda_1^{\left\langle N_j \right\rangle} - \lambda_2^{\langle N_j \rangle}}
\right).
\end{multline}

Note, that
\begin{equation}\label{eq:k-to-x}
  k = \left\lfloor \frac{x + \frac{l}{2}}{l_k}\right\rfloor = \left\lfloor \left(\frac{x}{l} + \frac{1}{2}\right)\left( N_j + 1 \right)\right\rfloor,
\end{equation}
where $x$ is the coordinate along a stick ($x=0$ corresponds to the stick's centre, see Fig.~\ref{fig:contacts}). When $\left\langle N_j \right\rangle \gg 1$, a continuous description is possible $i_k \to i(x)$. In this limiting case, the electrical conductivity is
\begin{equation}\label{eq:MFAsigma}
\sigma = \frac{ \langle N_j \rangle }{2 C  R_\text{s}} \left[ 1 - \sqrt{\frac{4 }{ \langle N_j \rangle \Delta} } \tanh\left(\sqrt{\frac{ \langle N_j \rangle \Delta}{4} }\right)\right].
\end{equation}

\subsubsection{Junction-resistance-dominated case}
Let $R_\text{s} =0$. The potentials of any two nearest contacts are equal
\begin{equation}\label{eq:JDRuk}
u_k - u_{k-1} =0,
\end{equation}
while the electric current in two adjacent segments of the conductor changes due to leakage in the contact
\begin{equation}\label{eq:JDRik}
i_k - i_{k-1} + \frac{{u_k - V_k}}{R_\text{j}}=0.
\end{equation}
These two equations can be combined
$$
i_k - 3i_{k-1} + 3 i_{k-2} -i_{k-3} = 0.
$$
The solution of this linear recurrence with constant coefficients is
\begin{equation}\label{eq:ikNj}
  i_k(\alpha,N_j) = k ( N_j  - k )\frac{E l \cos \alpha}{2 (N_j + 1) R_\text{j}}.
\end{equation}
The electric current in the conductor, averaged over all segments, is
$$
\left\langle i(\alpha,N_j) \right\rangle = \frac{1}{N_j + 1}\sum_{k=0}^{N_j} i_k =
\frac{E l \cos \alpha}{2 (N_j + 1)^2 R_\text{j}}\sum_{k=0}^{N_j} k ( N_j  - k ).
$$
Calculating the sum leads to the formula
$$
\langle i(\alpha,N_j) \rangle =\frac{N_j (N_j-1)}{12 R_\text{j}(N_j + 1)}E l \cos \alpha.
$$
Since the number of contacts on the conductor obeys the Poisson distribution,\cite{Heitz2004NT,Yi2004JAP,Callaghan2016PCCP,Kumar2017JAP,Kim2018JAP} the average current in the conductor is
$$
\langle i(\alpha) \rangle = \frac{E l \cos \alpha}{12 R_\text{j}} \sum_{k=0}^{\infty}\frac{k(k-1) \left(C n l^2\right)^k }{(k + 1) k!} \mathrm{e}^{-C n l^2} = \frac{E l \cos \alpha}{12 R_\text{j}} D(n),
$$
where
$$
D(n) = C n l^2 - 2   - 2\frac {1 - {\rm e}^{-C n l^2} }{C n l^2}.
$$

Note, $D(n) \approx C n l^2$, when $C n l^2 \gg 1$ .

The number of sticks intersecting a line of length $L$ perpendicular to the field is $ n l L \cos \alpha $. Thus, the total electric current in the sticks of all orientations through a cross section of the system is
$$
i = \frac{1}{\pi} \int\limits_{-\pi/2}^{\pi/2} \frac{E n l^2 L D(n)}{12 R_\text{j}} \cos^2 \alpha\, d\alpha =  \frac{E n l^2 L }{24 R_\text{j}} D(n).
$$
Hence, the electrical conductivity of the system under consideration is
\begin{equation}\label{eq:MFAsigmaJDRdiscr}
  \sigma =  \frac{ n l^2 }{24 R_\text{j}} D(n).
\end{equation}
When $C n l^2 \gg 1$ , then
\begin{equation}\label{eq:MFAsigmaJDRcont}
\sigma = \frac{n^2 l^4 }{12 \pi R_\text{j}}.
\end{equation}

When the average number of contacts per conductor is large enough, taking into account~\eqref{eq:k-to-x},  a continuous description is possible, viz., $i_k \to i(x)$. Then~\eqref{eq:ikNj} transforms into
$$
i(x;N_j,\alpha) = \left(\frac{x}{l} + \frac{1}{2}\right) \left[ N_j  - (N_j + 1)  \left(\frac{x}{l} + \frac{1}{2}\right) \right]\frac{E l \cos \alpha}{2 R_\text{j} }.
$$
Performing derivations similar to the general case considered earlier, we obtain
$$
\langle i \rangle = \left( \frac{n l^2}{\pi}  - 1 \right)\frac{E n l^2 L}{12 R_\text{j}}.
$$

The angle-averaged leakage current in the $k$-th contact, $i_k^{(l)} = i_k - i_{k-1}$,
$$
\langle i_k^{(l)} \rangle = -\frac{E l(2 k -N_j -1)}{\pi R_\text{j}(N_j + 1)}
$$
in this limiting case transforms into
\begin{equation}\label{eq:leakage}
  \langle i^{(l)}(x) \rangle  = -\frac{C E}{R_\text{j}} x.
\end{equation}

\section{Results and discussion}\label{sec:results}
\begin{figure}[!hbtp]
  \centering
  \includegraphics[width=0.85\columnwidth]{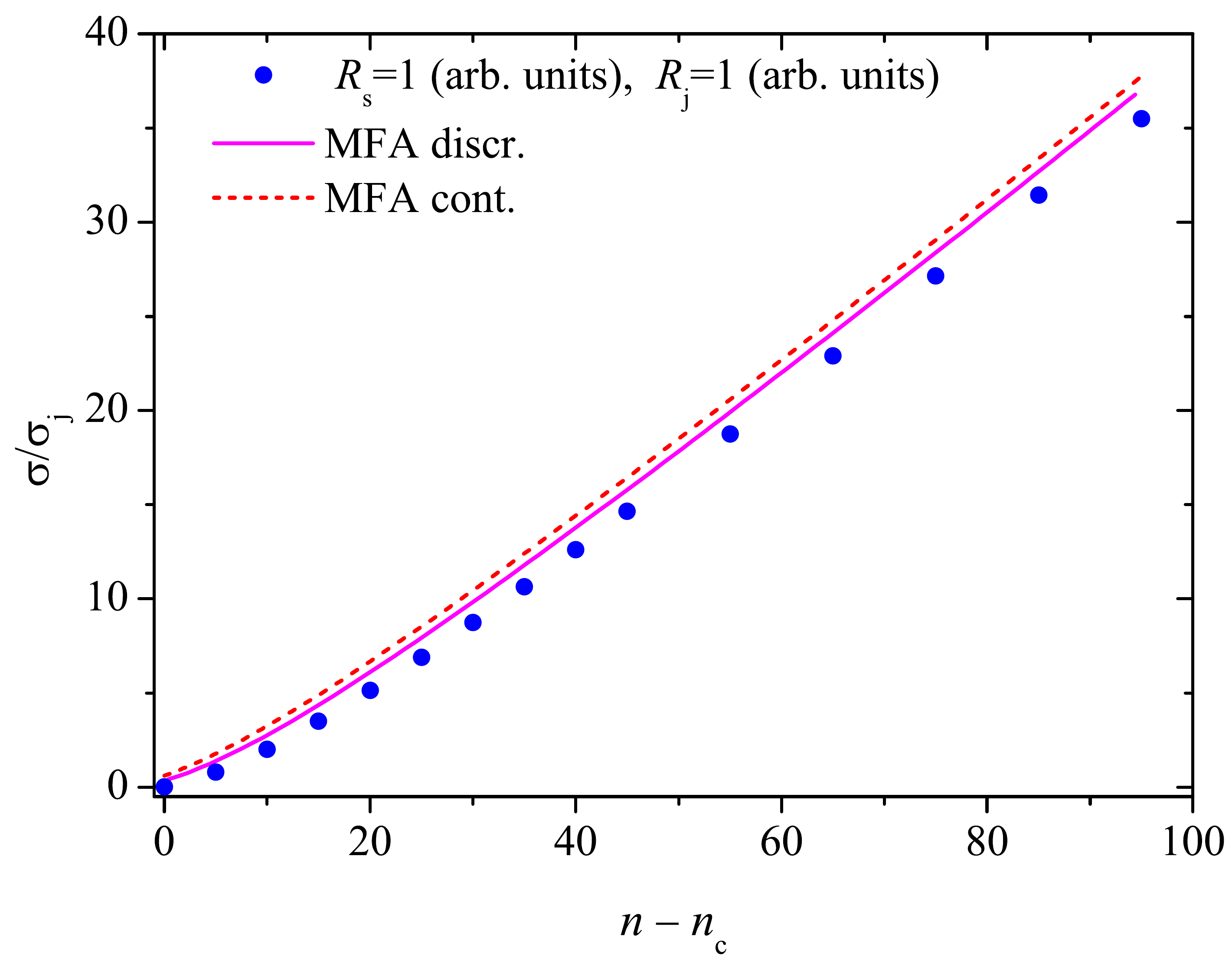}\\
  \caption{Comparison of the results of direct calculation of the electrical conductivity (circles) with the predictions of the MFA in the discrete approach (Eq.~\eqref{eq:MFAsigma-discr}, solid line) and in the continuum approach (Eq.~\eqref{eq:MFAsigma}, dashed line), $R_\text{s}=R_\text{j}$.\label{fig:compar}}
\end{figure}

Figure~\ref{fig:compar} compares the results of the direct calculation of electrical conductivity (circles) with the predictions of the MFA in the continuum approach (Eq.~\eqref{eq:MFAsigma}, dashed line) and in the discrete approach (Eq.~\eqref{eq:MFAsigma-discr}, solid line) for the case when the contact resistance is equal to the wire resistance, $R_\text{s}=R_\text{j}$.

Surprisingly, the prediction of the MFA is fairly close to the results of direct computation of the electrical conductivity even near the percolation threshold. As expected, the discrete approach (Eq.~\eqref{eq:MFAsigma-discr}) is better than the continuous approach (Eq.~\eqref{eq:MFAsigma}). However, the difference between these two predictions is insignificant.
Equation~\eqref{eq:MFAsigma} can therefore be used for most purposes because it is much simpler than Eq.~\eqref{eq:MFAsigma-discr}, while its accuracy is only marginally worse. There are at least two obvious reasons why the MFA slightly overestimates the electrical conductivity. First of all, the MFA completely ignores the existence of the percolation threshold. At the percolation threshold, only a small proportion of the conductive wires are involved in the electrical conductivity. By contrast, the MFA treats all wires as current-carrying. This leads to an incorrect starting point of the electrical conductivity curve. Secondly, the derivation of the master equation~\eqref{eq:MFAsigma-discr} is based on an assumption about evenly distributed contacts. Although, the derivation of the master equation may, in principle, be written without this assumption, unfortunately, we can hardly see a way to overcome some technical difficulties arising in this case.

\begin{figure}[!hbtp]
  \centering
  \includegraphics[width=0.85\columnwidth]{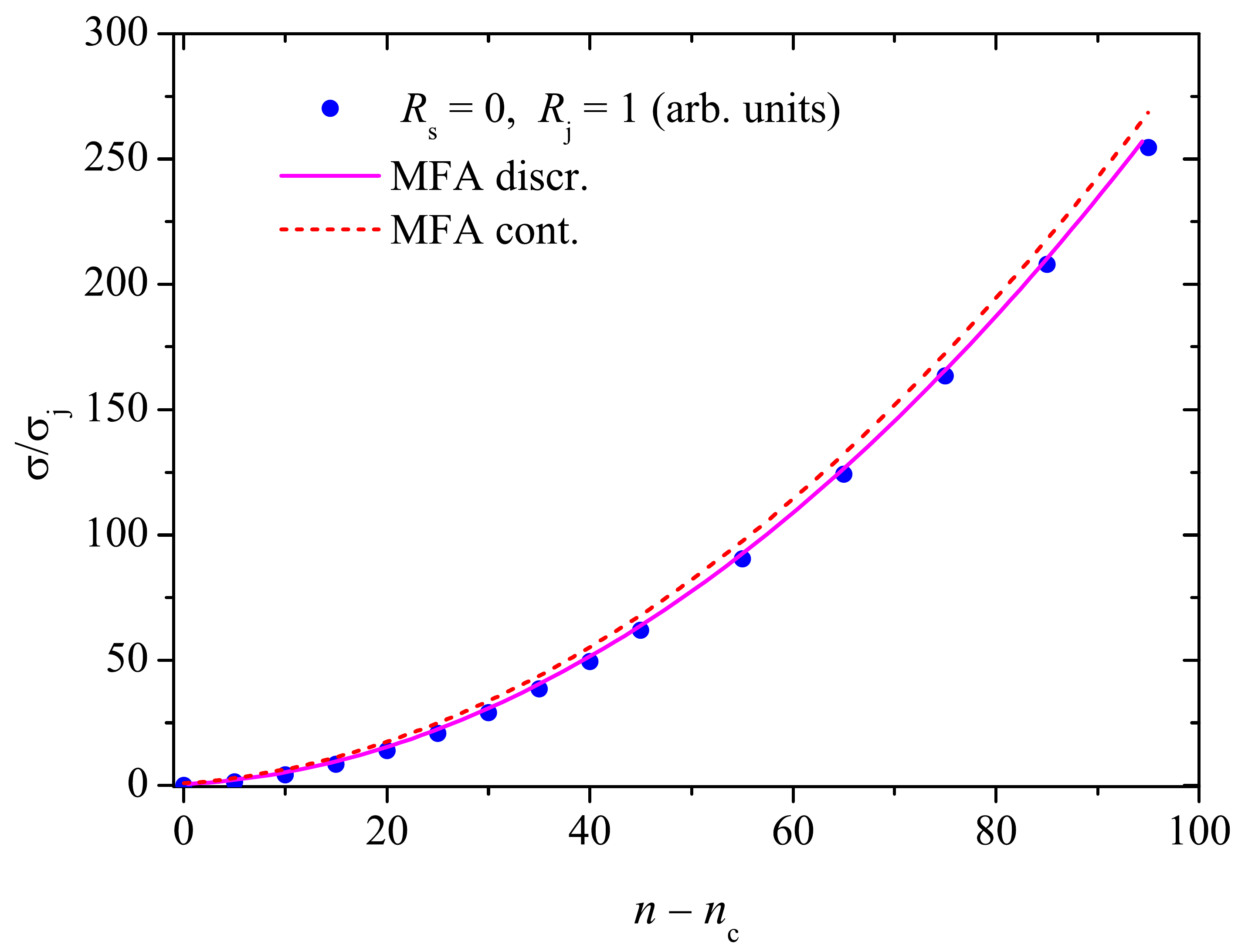}\\
  \caption{Comparison of the results of direct calculation of electrical conductivity (circles) using predictions of the mean field approximation in the discrete approach (Eq.~\eqref{eq:MFAsigmaJDRdiscr}, solid line) and in the continuous approach (Eq.~\eqref{eq:MFAsigmaJDRcont}, dashed line), $R_\text{j} \gg R_\text{s}$.\label{fig:JDRdiscr}}
\end{figure}

Figure~\ref{fig:JDRdiscr} compares the results of the direct electrical conductivity calculation (circles) with the predictions of the mean field approximation in the continuum approach (Eq.~\eqref{eq:MFAsigmaJDRcont}, dashed line) and discrete approach (Eq.~\eqref{eq:MFAsigmaJDRdiscr}, solid line) for the case when the contact resistance is much greater than the wire resistance, $R_\text{j} \gg R_\text{s}$. In this limiting case, which corresponds to unwelded wires, the prediction of the MFA is excellent in the discrete approach, and quite good in the continuous approach. Despite the prediction of Eq.~\eqref{eq:KumarR1},\cite{Kumar2017JAP} neither the MFA, nor the direct computations confirm that the electrical conductivity tends to a constant value as the number density of the wires increases.

\begin{figure}[!htb]
  \centering
  \includegraphics[width=0.85\columnwidth]{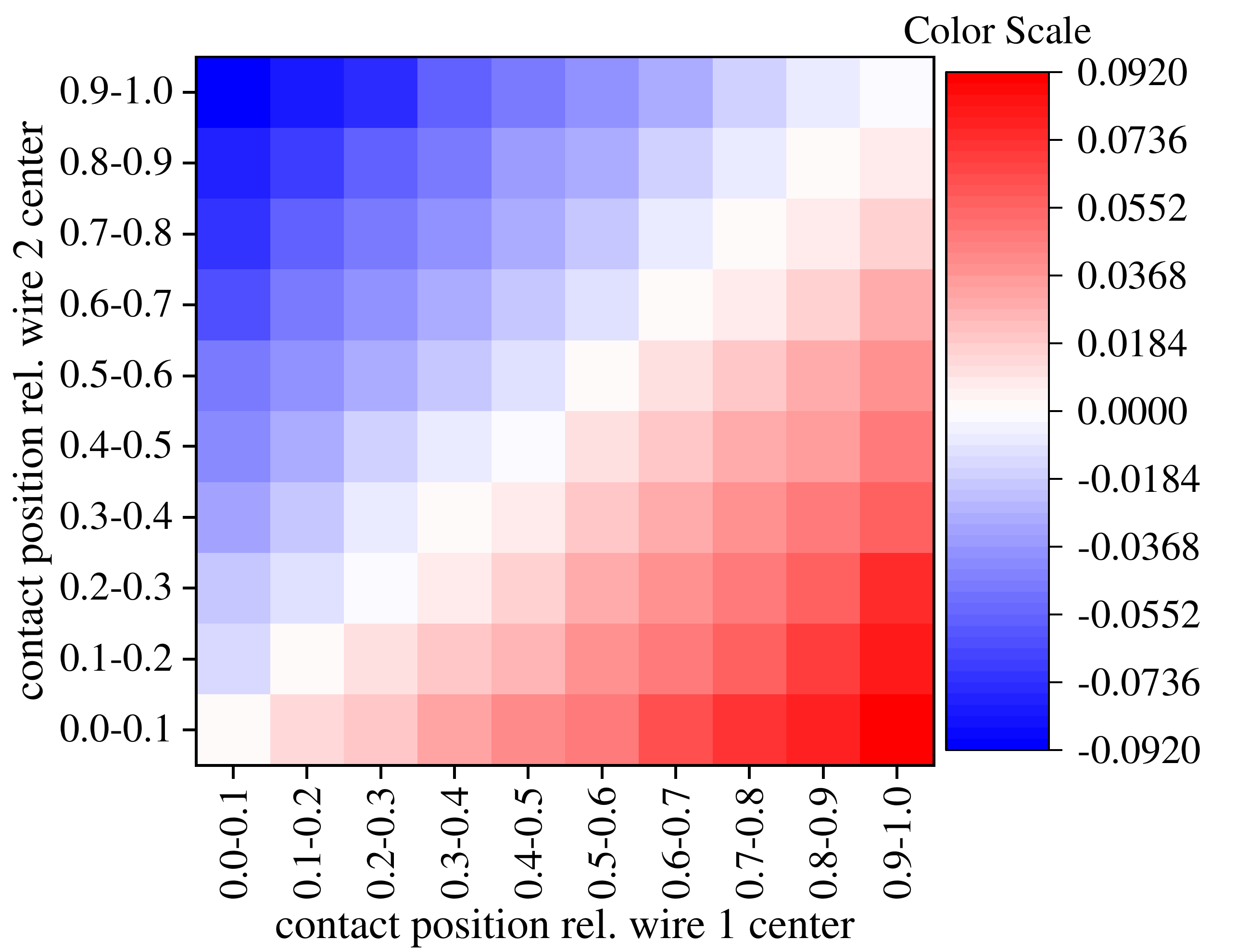}
  \caption{Example of the distribution of electric currents in contacts plotted against their positions on two intersecting conductors for the junction-resistance-dominated case ($R_\text{s}=0, R_\text{j} =1$); $n-n_c=50$.}\label{fig:ContactCurrentHM}
\end{figure}

Now, since the overall behaviour has become clear, we can delve into the details. Close up results of the computer simulations are presented in the next figures. The distribution of currents in the contacts is shown against their positions on two intersecting  conductors in Fig.~\ref{fig:ContactCurrentHM}.

The largest magnitudes can be observed when the contacts are located near the end of one conductor and close to the beginning of another. Both the magnitude and the direction of the electric current are accounted for. Although, our simulation confirms the previously published  results,\cite{Forro2018ACSN} they hardly support the speculation that, on average, only half of the contacts  carry current. On the contrary,  the computations evidenced, in our opinion, that all contacts are involved in the electrical conductivity.

\begin{figure}[!htb]
  \centering
  \includegraphics[width=0.85\columnwidth]{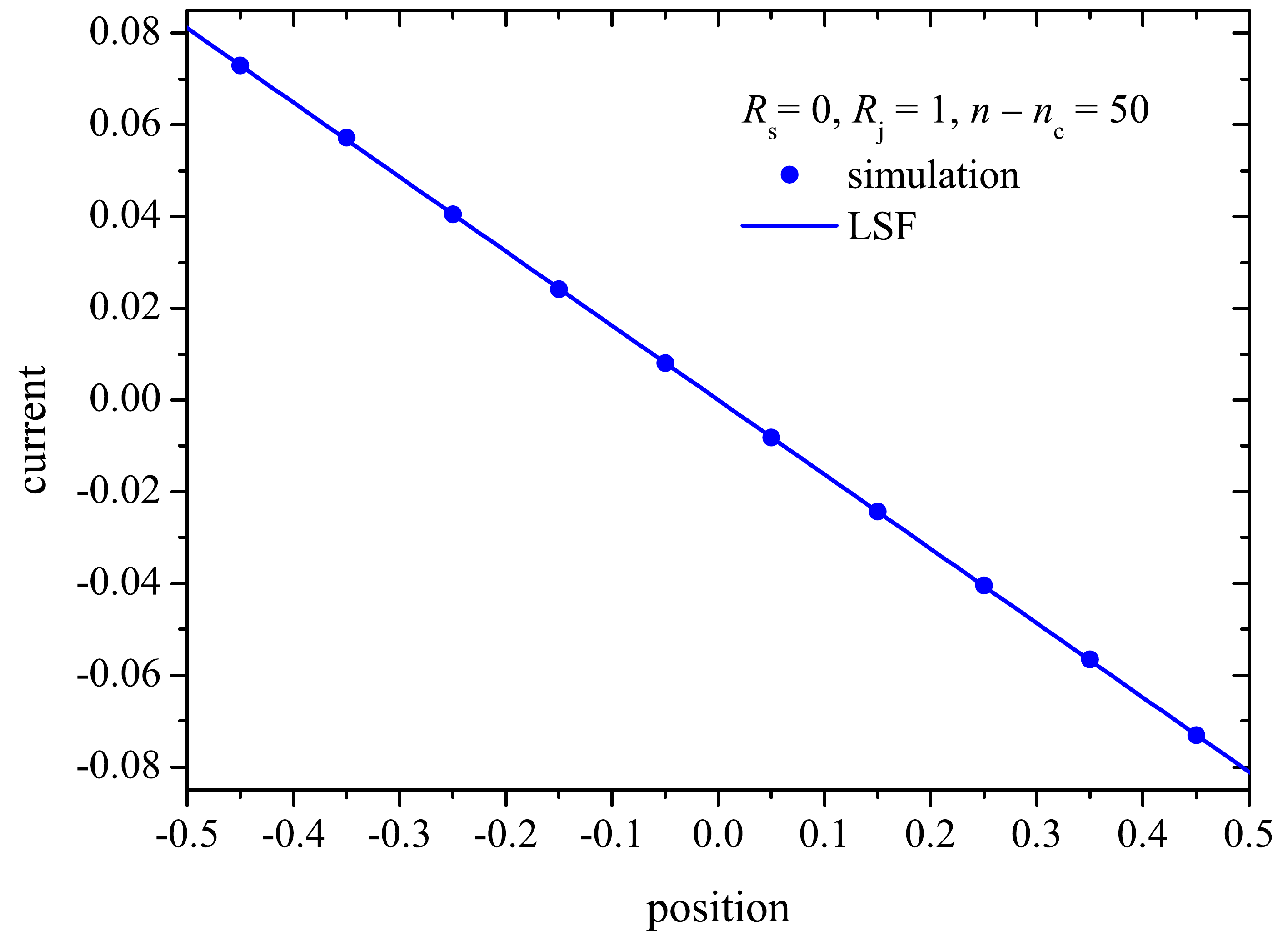}
  \caption{Example of the  distribution of currents in the contacts, shown relative to the position of the contacts  for the junction-resistance-dominated case ($R_\text{s}=0, R_\text{j} =1$); $n-n_c=50$.}\label{fig:JunctionCurrentsvsPosition}
\end{figure}

Figure~\ref{fig:JunctionCurrentsvsPosition} presents the average distribution of currents in the contacts against their positions on the conductor. Again, the result hardly supports introducing the concept of `the effective wire length'.\cite{Forro2018ACSN} The slope of the line depends on the number density of the conductors. However, the slope tends to the limiting value predicted by Eq.~\eqref{eq:leakage}.

\begin{figure}[!htb]
  \centering
  \includegraphics[width=0.85\columnwidth]{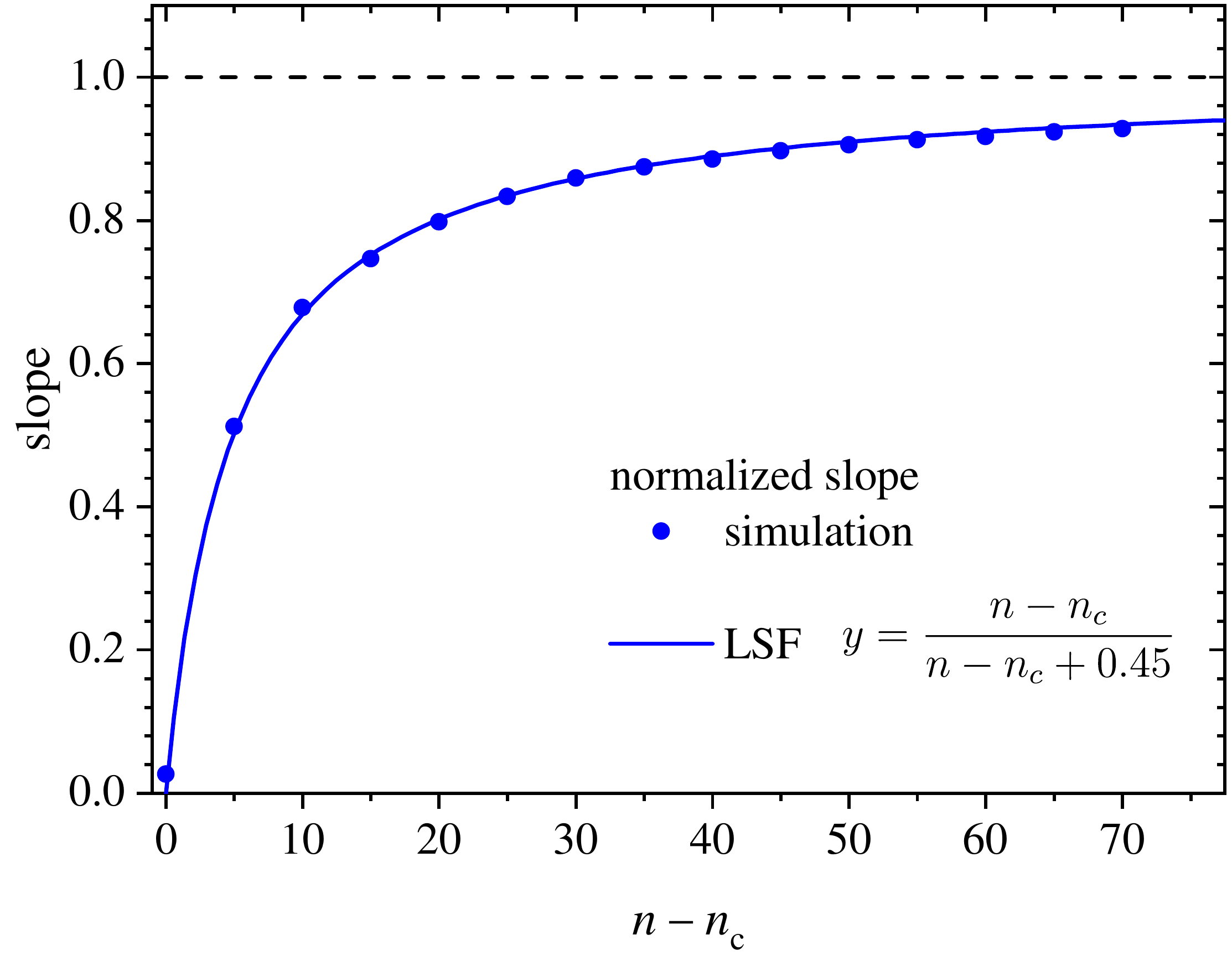}
  \caption{Normalized slope against the number density of the conductors for the junction-resistance-dominated case ($R_\text{s}=0, R_\text{j} =1$); $n-n_c=50$.}\label{fig:slopeJDR}
\end{figure}

Figure~\ref{fig:slopeJDR} presents the normalized (divided by $-C E /R_\text{j}$)  slope against the number density of the conductors. The dashed line corresponds to the theoretical value $C E /R_\text{j}$. This behaviour of the slope suggests that, despite the continuous MFA correctly describing the behaviour of the electrical conductivity, some details can be caught only when $n \gtrapprox 100$.

\begin{figure}[!htb]
  \centering
  \includegraphics[width=0.85\columnwidth]{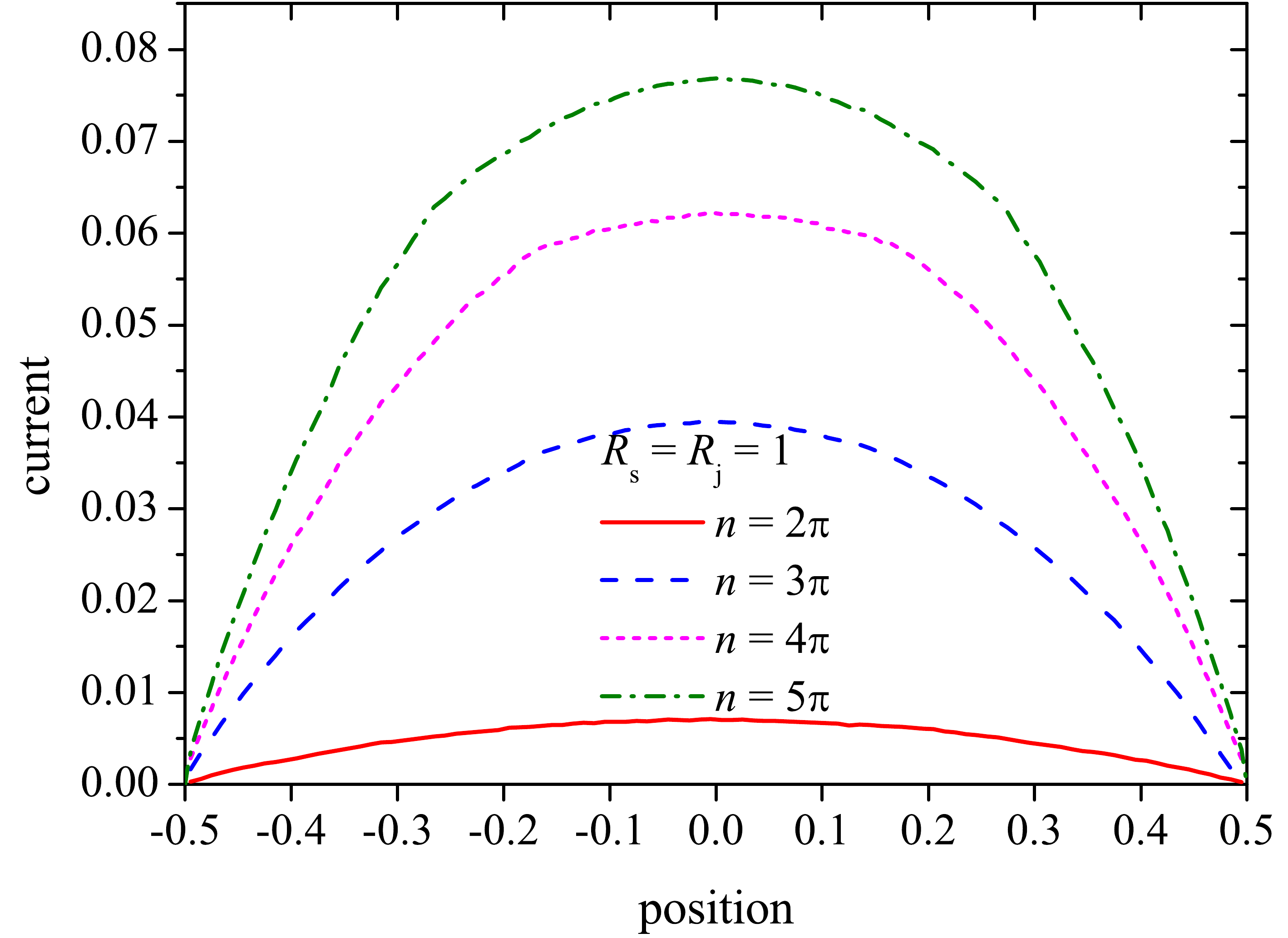}
  \caption{Average change in the electric current along the wire for different values of the number density of conductors when $R_\text{s}= R_\text{j} =1$.}\label{fig:CurrentsvsPositionJWR}
\end{figure}

Figure~\ref{fig:CurrentsvsPositionJWR} demonstrates the change in the electric current along the wire. This behaviour was obtained by averaging over all the conductors in the system under consideration. Computations have been performed for different values of the number density of the conductors when $R_\text{s}= R_\text{j} =1$.

\section{Conclusions}\label{sec:concl}
We have proposed an analytical model for the electrical conductivity in random nanowire networks. In our opinion, the underlying assumptions are reasonable and clear, while the derivation is simple. In contrast with previously proposed models,\cite{Zezelj2012PRB,Forro2018ACSN} our model contains neither adjustable parameters (like that in Eq.~\eqref{eq:RsheetZS}) nor correction factors (like `the effective wire length' in Eq.~\eqref{eq:Forro}). Unlike that model,\cite{Kumar2017JAP} ours demonstrates reasonable limiting behaviour when $R_\text{j} \gg R_\text{s}$. Generalisations of the model to wires with a length distribution,\cite{Hicks2009,Borchert2015} to waved or bent wires\cite{Yi2004JAP,Yin2017,Hicks2018JAP} as well to systems of ordered or aligned linear wires\cite{Ackermann2016SR,Cho2017,Yin2019,Wan2020,Meng2021} are straightforward.

Although recent studies have shown that the properties of networks formed by rods differ significantly if the rods have zero thickness (2D networks) as opposed to finite thickness (quas-3D networks),\cite{Daniels2021}  in real nanowire systems, this effect is, presumably, not very  significant since the wires are flexible. Moreover, welding may ensure additional contacts between metallic nanowires.\cite{Park2016}  However, even for quasi-3D systems, our model may be generalized by accounting for the number of contacts between the wires in such stacked system.

\section*{Author Contributions}
Y.Y.T. contributed in supervision, conceptualization, methodology, formal analysis, and writing (original draft).
I.V.V. contributed in formal analysis, validation, and writing (review \& editing).
A.V.E. contributed in software and investigation.

\section*{Conflicts of interest}
There are no conflicts to declare.

\section*{Acknowledgements}
Y.Y.T. and A.V.E. acknowledge the funding from the Foundation for the Advancement of Theoretical Physics and Mathematics ``BASIS'', grant~20-1-1-8-1.



\balance


\bibliography{MFAdiscr} 
\bibliographystyle{rsc} 

\end{document}